\documentclass[prc,superscriptaddress,twocolumn,showpacs]{revtex4}
\usepackage{graphicx,color,amsmath,amssymb}
\usepackage{mathrsfs}
\usepackage{multirow}
\usepackage{bm}

\def\mbold#1{\mbox{\boldmath $#1$}}
\def\non{\nonumber}

\begin{document}

% Use the \preprint command to place your local institutional report
% number in the upper righthand corner of the title page in preprint mode.
% Multiple \preprint commands are allowed.
% Use the 'preprintnumbers' class option to override journal defaults
% to display numbers if necessary
%\preprint{}

%Title of paper
\title{Two neutron decay from the $2_1^+$ state of $^6$He}

% repeat the \author .. \affiliation  etc. as needed
% \email, \thanks, \homepage, \altaffiliation all apply to the current
% author. Explanatory text should go in the []'s, actual e-mail
% address or url should go in the {}'s for \email and \homepage.
% Please use the appropriate macro foreach each type of information

% \affiliation command applies to all authors since the last
% \affiliation command. The \affiliation command should follow the
% other information
% \affiliation can be followed by \email, \homepage, \thanks as well.

\author{Yuma Kikuchi}
\email[Electronic address: ]{yuma@rcnp.osaka-u.ac.jp}
\affiliation{Research Center for Nuclear Physics, Osaka University, Ibaraki, Osaka 567-0047, Japan}

\author{Takuma Matsumoto}
\affiliation{Department of Physics, Kyushu University, Fukuoka 812-8581, Japan}

\author{Kosho Minomo}
\affiliation{Department of Physics, Kyushu University, Fukuoka 812-8581, Japan}

\author{Kazuyuki Ogata}
\affiliation{Research Center for Nuclear Physics, Osaka University, Ibaraki, Osaka 567-0047, Japan}

%Collaboration name if desired (requires use of superscriptaddress
%option in \documentclass). \noaffiliation is required (may also be
%used with the \author command).
%\collaboration can be followed by \email, \homepage, \thanks as well.
%\collaboration{}
%\noaffiliation

\date{\today}

\begin{abstract}
Decay mode of the $2_1^+$ resonant state of $^6$He populated by
the $^6$He breakup reaction by $^{12}$C at 240~MeV/nucleon is investigated.
The continuum-discretized coupled-channels method is adopted to
describe the formation of the $2_1^+$ state, whereas its decay is
described by the
complex-scaled solutions of the Lippmann-Schwinger equation.
From analysis of invariant mass spectra with respect to the $\alpha$-$n$ and
$n$-$n$ subsystems, coexistence of two decay modes
is found. One is the simultaneous decay of two neutrons correlating with each
other and the other is the emission of two neutrons to the opposite directions.
The latter is found to be free from the final state interaction and
suggests existence of a di-neutron in the $2_1^+$ state of $^6$He.
\end{abstract}

% insert suggested PACS numbers in braces on next line
% insert suggested keywords - APS authors don't need to do this
%\keywords{}

%\maketitle must follow title, authors, abstract, \pacs, and \keywords
\maketitle

% body of paper here - Use proper section commands
% References should be done using the \cite, \ref, and \label commands
% Put \label in argument of \section for cross-referencing
%\section{\label{}}

{\it Introduction.}
Elucidation of unbound nuclei outside the drip line~\cite{Lun12}
as well as unbound excited (resonant) states of unstable nuclei~\cite{Tsh12}
is a hot subject in nuclear physics.
These particle-unbound states have been investigated by means of decay-particle measurements.
Clarification of their decay mode is crucial to extract structural information on
the unbound states. Furthermore, the decay mode itself is an interesting subject
for understanding of the dynamical properties of them.
If we consider a particle-unbound state that decays into three particles, we have several
possibilities of the decay mode. For instance, in the decay process of the $2_1^+$
state of $^6$He into $\alpha$ and two neutrons,
the following decay modes can be considered.
1) Emission of one neutron is followed by that of the other from the $3/2^-$
resonance (ground state) of $^5$He; below we call this state just $^5$He.
2) Two neutrons are emitted simultaneously (not via $^5$He), with correlating
with each other, i.e., preferring the low two-neutron ($2n$) relative energy.
3) Same as 2) but two neutrons are emitted independently.
These decay modes have been intensively discussed for two-proton ($2p$) decay
phenomena~\cite{Ego12,Cha10,Boc89}.
In the true $2p$ decay, by definition, 1) is forbidden because of the condition that the
ground state of a $2p$-decay nucleus is located below the (core$+p$)-$p$ threshold energy, where (core$+p$) is a two-body resonant state.
If one considers, however, an excited state of a $2p$-decay nucleus, 1) also may happen in general.
Following the terminology used in preceding studies, we call 1), 2), and
3) the sequential decay, the di-neutron decay, and the democratic decay, respectively.
Although sometimes the democratic decay means both 2) and 3), we differentiate 2) from 3)
in this study.
Clarification of the decay mode of three-body systems will be a fascinating subject
in view of nontrivial dynamics of three-body decaying systems.
To achieve this, we need a framework that describes dynamics of both the formation and decay
of particle-unbound states.

In Ref.~\cite{Kik09}, the method of the complex-scaled solutions of the Lippmann-Schwinger equation
(CSLS) was developed. In CSLS the Lippmann-Schwinger formalism is combined with the complex
scaling method (CSM)~\cite{Agu71}; CSLS allows an accurate description of decay of many-body systems.
Then CSLS was applied to the Coulomb breakup of $^6$He by $^{208}$Pb at 250~MeV/nucleon to study
a possible correspondence between breakup observables and the di-neutron correlation in
the ground state of $^6$He~\cite{Kik10}.
This work can also be interpreted as an investigation of the decay mode of the nonresonant
$1^-$ state of $^6$He excited by $^{208}$Pb. The main conclusion of Ref.~\cite{Kik10} was the
dominance of the sequential decay through $^5$He governed by
the final state interaction (FSI).
Another important finding was a peak in the $2n$ invariant
mass spectrum. Since this peak is located around the $2n$ virtual state energy,
it suggests the di-neutron decay. It should be noted,
however, that the peak was found to be generated by a rearrangement process due to the FSI
from the $^5$He$+n$ configuration to the $\alpha+2n$ in the decay process.
This indicates that even the decay mode changes, from the sequential decay to
the di-neutron decay in this case. Thus, the FSI is found to play a crucial role in
the decay of the $1^-$ nonresonant state of $^6$He.

A possible shortcoming of the work of Ref.~\cite{Kik10} is that the breakup process by $^{208}$Pb
was simplified by a one-step electric dipole (E1) transition. Although this simplification
is expected to work quite well for breakup processes by $^{208}$Pb at intermediate energies,
it severely restricts the applicability of CSLS. To study the decay mode of a resonant
state, more realistic description of the formation of the resonance will be necessary.
In this work, we extend the CSLS analysis in Ref.~\cite{Kik10} by incorporating
a sophisticated reaction model to describe the formation of particle-unbound states.
We focus on the $2_1^+$ resonance of $^6$He and aim at clarifying its decay mode.
We consider $^6$He breakup reaction by $^{12}$C at 240~MeV/nucleon
as a formation process of the $2_1^+$ $^6$He resonance. We describe this process by
means of the continuum-discretized coupled-channels method (CDCC)~\cite{Kam86,Aus87,Yah12}.
CDCC is a non-perturbative quantum-mechanical model that has been successfully
applied to various breakup reactions in a wide range of incident energies.
Theoretical foundation of CDCC is given in Refs.~\cite{Aus89,Aus96,Yah12}.
The decay of the $2_1^+$ resonance into $\alpha$ and two neutrons
is then investigated by CSLS. This combination of CDCC and CSLS, {\it CDCC-CSLS},
is a powerful method to describe formation and decay of particle-unbound
states in a unified manner.

{\it Formalism.}
We assume that the $^6$He$+{}^{12}$C scattering is described as
an $\alpha+n+n+{}^{12}$C four-body system.
The total Hamiltonian is defined by
\begin{eqnarray}
 H&=&K_R +U+H_6
\end{eqnarray}
with
\begin{eqnarray}
 U&=&U_n^{\rm Nucl} + U_n^{\rm Nucl} + U_\alpha^{\rm Nucl}
  + U_\alpha^{\rm Coul},\\
 H_6&=&K_y+K_r
  +v_{nn}+v_{\alpha n}+v_{\alpha n},
\end{eqnarray}
where $H_{6}$ is an internal Hamiltonian of $^6$He.
The relative coordinate between $^6$He and $^{12}$C is denoted by
$\bm{R}$, and  the internal coordinates of $^6$He are denoted by
a set of Jacobi coordinates $\bm{\xi}=(\bm{y},\bm{r})$.
The kinetic energy operator and the reduced mass associated with the coordinate $\bm{s}$
is represented by $K_{s}$ and $\mu_s$, respectively.
The $n$-$n$ ($\alpha$-$n$) interaction is denoted by
$v_{nn}$ ($v_{\alpha n}$), and $U_{x}^{\rm Nucl}$ and
$U_{x}^{\rm Coul}$ are the nuclear and Coulomb potential between $x$ and $^{12}$C, respectively.

In CDCC with the pseudostate discretization method~\cite{Mor01, Mat03,
Ega04},
the scattering is assumed to take place in the model space ${\cal P}$
defined by
\begin{eqnarray}
 {\cal P}=\sum_{i}|\Phi_{i}\rangle \langle \Phi_{i}|,
 \label{eq:com-set}
\end{eqnarray}
where ${\Phi}_{i}$ is the $i$th eigenstate obtained by
diagonalizing $H_{6}$ with $L^2$-type basis functions.
The four-body Schr\"odinger equation is then solved
in the model space:
\begin{align}
{\cal P}(H-E_{\rm tot}){\cal P}|\Psi^{(+)}_{\rm CDCC} \rangle =0.
\label{eq:4b-Schr}
\end{align}
The model space assumption has already been justified by the fact that
calculated elastic and breakup cross sections converge with respect to
extending the model space.
Details of how to solve the four-body CDCC equation~\eqref{eq:4b-Schr} are
shown in Refs.~\cite{Mat04,Mat06,Rod08}.

The CDCC $T$-matrix element to the $i$th discrete breakup state $\Phi_i$
with an eigenenergy $\varepsilon_i$ is given by
\begin{eqnarray}
 T_i^{\rm CDCC}&=&
  \langle\Phi_i\chi_i^{(-)}(\mbold{P}_i)|U-U_{^6\rm He}^{\rm Coul}
  |\Psi_{\rm CDCC}^{(+)}\rangle,
\end{eqnarray}
where $U_{^6{\rm He}}^{\rm Coul}$ is the Coulomb interaction between $^6$He
and $^{12}$C. The final-state wave function $\chi_i^{(-)}$ with the
incoming boundary condition for the relative motion regarding
$\mbold{R}$ is defined by
\begin{eqnarray}
 \left[
  K_R
  +U_{^6\rm He}^{\rm Coul}
  -(E_{\rm tot}-\varepsilon_i)
 \right]|\chi_i^{(-)}(\mbold{P}_i)\rangle&=&0,
\end{eqnarray}
where the asymptotic relative momentum $\mbold{P}_i$ (in the unit of $\hbar$) for $\Phi_i$
satisfies $E_{\rm tot}-\varepsilon_i=(\hbar^2P_i^2)/(2\mu_R)$.
Using the CDCC $T$-matrix element, the exact $T$-matrix element to a
continuum breakup state is well approximated by
\begin{eqnarray}
T_{\varepsilon} (\mbold{p},\mbold{k},\mbold{P})
&\approx& \sum_i \langle \Phi^{(-)}_\varepsilon(\mbold{p},\mbold{k}) | \Phi_i
\rangle
T^\text{CDCC}_i\non\\
&=& \sum_i f_i(\mbold{p},\mbold{k}) T^\text{CDCC}_i,
\end{eqnarray}
where the momenta $\mbold{p}$, $\mbold{k}$, and $\mbold{P}$ are the
asymptotic relative momenta regarding the coordinate $\mbold{y}$,
$\mbold{r}$, and $\mbold{R}$, respectively, and $\Phi^{(-)}_\varepsilon$ is
the exact three-body continuum wave function of $^6$He with the total energy
$\varepsilon=(\hbar^2p^2)/(2\mu_y)+(\hbar^2k^2)/(2\mu_r)$. To obtain the
smoothing function
$f_i(\mbold{p},\mbold{k})$, we use CSLS that describes the three-body
scattering states with correct boundary conditions:
\begin{eqnarray}
f_i (\mbold{p},\mbold{k}) &=& \langle \varphi_0 (\mbold{p},\mbold{k}) |
 \Phi_i \rangle+\sum_n\hspace{-0.46cm}\int
 \langle \varphi_0(\mbold{p},\mbold{k}) |
 \hat{V} U^{-1}(\theta) | \Phi^\theta_n \rangle
 \non\\
 &&{}\times\frac{1}{\varepsilon-\varepsilon^\theta_n}
 \langle \tilde\Phi^\theta_n | U(\theta) | \Phi_i \rangle,
\end{eqnarray}
where $\hat{V}=v_{nn}+v_{\alpha n} + v_{\alpha n}$ and $U(\theta)$ is the
complex-scaling operator. $\varphi_0$ is a three-body plane wave with a
set of relative momenta $(\mbold{p},\mbold{k})$, and $\Phi_n^\theta$ and
$\varepsilon_n^\theta$ are the $n$th eigenstate and its eigenenergy, which are
obtained by diagonalizing the complex-scaled $^6$He Hamiltonian,
$H_6^\theta$, by $L^2$-type basis functions.
In the present calculation, we adopt the Gaussian expansion
method (GEM)~\cite{Hiy03} to obtain $\Phi_i$ and $\Phi_n^\theta$.

To investigate the decay mode of the $2_1^+$ state of $^6$He, we
calculate the double differential cross sections with respect to the relative energies,
$\varepsilon_1$ and $\varepsilon_2$, of binary subsystems:
\begin{eqnarray}
\frac{d^2\sigma}{d\varepsilon_1 d\varepsilon_2}
&=& \frac{(2\pi)^4\mu_R}{\hbar^2 P_0}
\int d\mbold{p} d\mbold{k} d\mbold{P}
\left|
 T_{\varepsilon}
 (\mbold{p},\mbold{k},\mbold{P})\right|^2\non\\
&&\times\delta \left(E_{\rm tot} - \frac{\hbar^2P^2}{2\mu_R}
                   - \varepsilon_1 - \varepsilon_2 \right)\non\\
&&\times\delta \left(\varepsilon_1 - \frac{\hbar^2k^2}{2\mu_r}\right)
        \delta\left(\varepsilon_2 - \frac{\hbar^2p^2}{2\mu_y}\right),
\label{eq:Xsec}
\end{eqnarray}
where $P_0$ is the incident momentum of $^6$He in the center-of-mass (c.m.) system.

{\it Numerical input.}
As for the interactions $v_{nn}$ and $v_{\alpha n}$ in $H_6$, we take the
Minnesota~\cite{Tho77} and the KKNN~\cite{Kan79} potential, respectively.
We increase the depth of $v_{\alpha n}$ by 3\% to reproduce the ground state
energy of $^6$He.
The antisymmetrization between a valence neutron and a neutron
in $\alpha$ is treated approximately with the orthogonality condition
model~\cite{Sai69}.

In GEM, we take the Gaussian range parameters $r_i$ ($i=1,$ 2, ..., $N$)
that lie in geometric progression.
In the diagonalization of $H_6$, we use $(N$, $r_1$, $r_N$)=(10, 0.5~fm, 10~fm)
for each of the Jacobi coordinates; for the coordinate between two neutrons
we take $r_1=0.1$~fm. In the diagonalization of $H_6^\theta$,
we adopt $(N$, $r_1$, $r_N$)=(20, 0.2~fm, 40~fm). It is known that
finer and wider bases are required to properly
describe the many-body resonant and continuum solutions simultaneously in
the CSM~\cite{Aoy06}.

We include 66, 82, and 100 eigenstates of $H_6$ for the $0^+$, $1^-$, and $2^+$
states of $^6$He, respectively, in the CDCC calculation.
These states are located below $\varepsilon=30$~MeV.
We solve Eq.~\eqref{eq:4b-Schr} by means of eikonal CDCC~\cite{Yah12,Oga03,Oga06}
up to $R=30$~fm. The distorting potentials between the constituents of $^6$He
and $^{12}$C are calculated by a microscopic folding model. Nuclear densities
of $\alpha$ and $^{12}$C are obtained by Hartree-Fock calculation with the Gogny D1S
force~\cite{Ber91}. As for the effective nucleon-nucleon interaction, we adopt
the Melbourne g-matrix~\cite{Amo00}.
The present calculation has no free adjustable parameters.
The numerical results shown below are converged with the model space described above.

\begin{figure}[htb]
\includegraphics[width=8cm,clip]{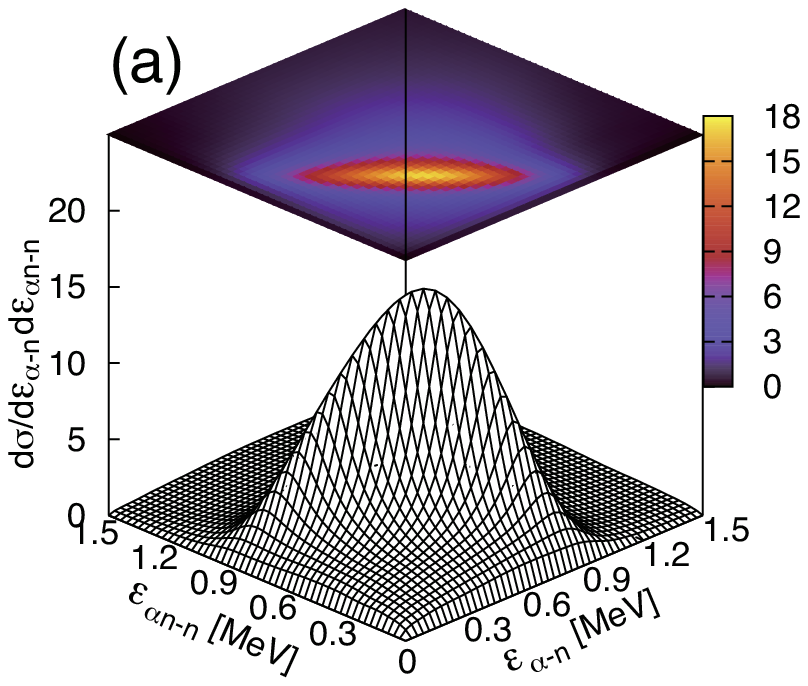}
\\
\vspace{0.5cm}
\includegraphics[width=8cm,clip]{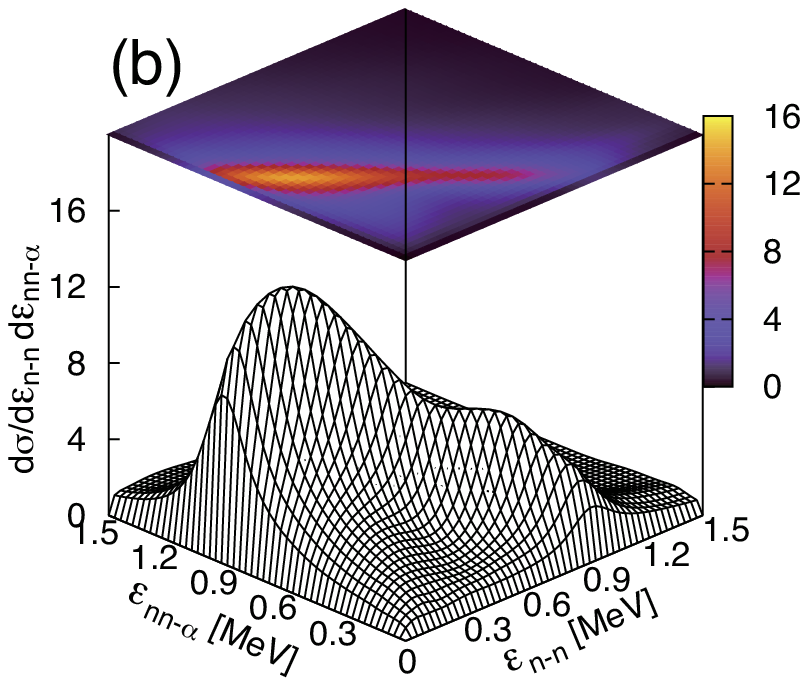}
\caption{\label{fig:2D_full}
(Color online) DDBUX of $^6$He by $^{12}$C at 240~MeV/nucleon.
Panels (a) and (b) represent the cross sections with respect to the subsystem energies of $\alpha$-$n$ and $n$-$n$, respectively (see the text for detail).}
\end{figure}

{\it Results.}
We show in Fig.~\ref{fig:2D_full} the double-differential
breakup cross section (DDBUX) calculated by CDCC-CSLS.
In panel (a), we show the DDBUX with respect to
the energy between $\alpha$ and a neutron ($\varepsilon_{\alpha\mbox{-}n}$),
and that between the other neutron and the c.m. of the $\alpha$-$n$
system ($\varepsilon_{\alpha n\mbox{-}n}$).
Similarly, the DDBUX with respect to the $2n$ relative energy ($\varepsilon_{n\mbox{-}n}$)
and the energy between the c.m. of the $2n$ system and $\alpha$ ($\varepsilon_{nn\mbox{-}\alpha}$)
is shown in panel (b).
One clearly sees the ridge structures in both panels corresponding to the
total energy $\varepsilon$ of the $\alpha+n+n$ three-body system around 1.0~MeV.
This structure comes from the
$2_1^+$ resonance of $^6$He obtained at 0.98~MeV above the three-body threshold
with the decay width of 0.27~MeV in the present calculation.
This clear {\lq\lq}observation'' of the $2_1^+$ resonance is an important
feature of the breakup process, or inelastic scattering in a wide sense,
by $^{12}$C mainly due to nuclear interactions.

\begin{figure}[htb]
\includegraphics[width=8cm,clip]{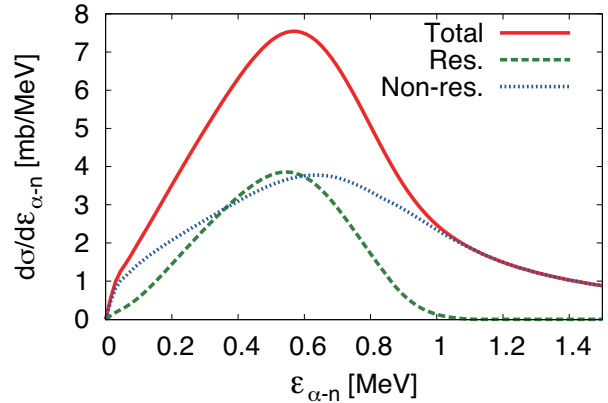}
\caption{\label{fig:inv1}
(Color online) Invariant mass spectra with respect to $\varepsilon_{\alpha\mbox{-}n}$ (solid line).
The dashed and dotted lines show the resonant and nonresonant parts, respectively.}
\end{figure}
Next we show the invariant mass spectrum and discuss the decay mode of the
$\alpha+n+n$ system, that of the $2_1^+$ resonant state in particular.
The solid line in Fig.~\ref{fig:inv1} shows $d\sigma/d\varepsilon_{\alpha\mbox{-}n}$
and the dashed line shows its resonant part extracted by gating $\varepsilon$ within the
range of the energy of the $2_1^+$ state, i.e., $\varepsilon=0.98 \pm 0.27/2$~MeV.
The rest shown by the dotted line is interpreted as the nonresonant part of
$d\sigma/d\varepsilon_{\alpha\mbox{-}n}$.
For the nonresonant part, one sees the peak around 0.7~MeV corresponding to
the energy of $^5$He, which is the same as in the spectrum
for the Coulomb breakup~\cite{Kik10}. Thus, one can conclude that
for the nonresonant decay, i.e., for the decay from nonresonant continuum states
of the $\alpha+n+n$ system, the sequential decay is dominant.

The property of the resonant part is quite different from this. A peak is found
around 0.5~MeV, about half the total energy $\varepsilon$ of the $2_1^+$ state, which is
somewhat lower than the energy of $^5$He.
Therefore, one can find that the sequential decay is suppressed in this case.
Instead, the result (dotted line) indicates that two neutrons are
emitted with equally sharing
the total energy of the three-body system. Thus, the di-neutron
decay or the democratic decay or both is suggested.

This result can be explained as follows. When $^5$He is formed under the
condition of $\varepsilon\sim 1.0$~MeV, the other neutron has very low energy below about 200~keV.
Since the (p3/2)$^2$ configuration is dominant in the $2_1^+$ state of $^6$He,
the second neutron has a centrifugal barrier with respect to $\alpha$.
Therefore, the second neutron hardly penetrates the barrier, which results in
the suppression of the $^5$He$+n$ configuration, hence the sequential decay,
in the decay of the $2_1^+$ state of $^6$He. It should be noted that for the Coulomb breakup,
in which the decay from the $1^-$ state is dominant, the second neutron
can be an s-wave. Thus, the suppression of the sequential decay is not the
case. Another remark is that for the nonresonant decay (dashed line) there is no
restriction on $\varepsilon$, which also allows the sequential decay.

\begin{figure}[htb]
\includegraphics[width=8cm,clip]{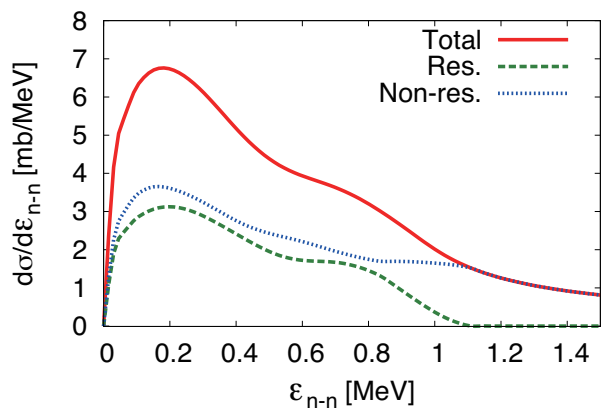}
\caption{\label{fig:inv2}
(Color online) Same as Fig.~\ref{fig:inv1} but for $d\sigma/d\varepsilon_{n\mbox{-}n}$.}
\end{figure}

To pin down the decay mode of the two neutrons, next we discuss
$d\sigma/d\varepsilon_{n\mbox{-}n}$ shown in Fig.~\ref{fig:inv2}.
Two peaks are found in the spectrum. The first peak around 0.2~MeV suggests
the di-neutron decay, as in the previous study of the Coulomb breakup~\cite{Kik10}.
It should be noted that this di-neutron decay does not mean the direct emission
of a di-neutron existed in the $2_1^+$ state. As in the Coulomb breakup
case, this di-neutron is found to be formed by the FSI in the decay process.
The two neutrons are emitted with correlating with each other, indicating
that the two have the same energy and are emitted to the same direction.
On the other hand, the second peak (or shoulder) around 0.8~MeV, which turns out to
come from the resonant part, suggests a completely different decay mode.
The $2n$ relative energy at the second peak, 0.8~MeV, almost exhausts the total
energy of the $2_1^+$ state ($\sim 1.0$~MeV). Therefore, it suggests that two neutrons
are emitted to the opposite directions with equally sharing the total energy,
i.e., the democratic decay is realized.

The democratic decay of the $2_1^+$ state is an important finding of
the present study. Because of the kinematics of the two neutrons in this
mode, it is expected that the FSI plays no important role.
In fact, the $n$-$n$ interaction favors the virtual state of the $2n$ system,
which corresponds to the first peak in $d\sigma/d\varepsilon_{n\mbox{-}n}$. Furthermore,
the $\alpha$-$n$ interaction that favors the sequential decay is also not important
in the decay of the $2_1^+$ state as shown in Fig.~\ref{fig:inv1}.
Therefore, one can conclude that the second peak in $d\sigma/d\varepsilon_{n\mbox{-}n}$ is free from
the FSI and directly reflects the structural property of the $2_1^+$ state of $^6$He.
The di-neutron correlation in a many-body system is characterized by the spatial
correlation between the two neutrons, which indicates a relatively high momentum
between them. This is consistent with the kinematics corresponding to the democratic
decay found in the present study. Thus, the second peak in $d\sigma/d\varepsilon_{n\mbox{-}n}$ is
possible evidence of a di-neutron in the $2_1^+$ state of $^6$He.
It should be noted that the existence of the $2_1^+$ state is, obviously, due to the FSI.
What clarified above is that the FSI plays no essential role during the democratic decay,
i.e., after the formation of the $2_1^+$ state.

{\it Summary.}
We investigated the decay mode of the $2_1^+$ resonant state of $^6$He formed by
the breakup of $^6$He by $^{12}$C at 240~MeV/nucleon. The formation and decay processes
were described by CDCC and CSLS, respectively. A clear ridge structure was found in
the DDBUX corresponding to the $2_1^+$ resonant energy. Analysis of the invariant mass
spectra showed that the sequential decay of the $2_1^+$ state through $^5$He was
suppressed because of
the small phase space, in contrast to the conclusion of the previous analysis of the
Coulomb breakup process.
Instead, we found the coexistence of the di-neutron decay and the democratic decay.
The former is due to the FSI, while the latter is free from the FSI. The democratic decay
is possible evidence of the existence of a di-neutron in the $2_1^+$ state of $^6$He.

This research was supported in part by Grant-in-Aid of the Japan
Society for the Promotion of Science (JSPS).

\end{document}